\documentclass[usegraphicx,a4paper]{mn2e}
\usepackage[totalwidth=500pt,totalheight=680pt]{geometry}
\usepackage{times}
\newlength{\colwidth}
\title[HST imaging of ULAS~J1120+0641]%
{No excess of bright galaxies around the redshift 7.1 quasar ULAS~J1120+0641}
\author[C.~Simpson et al.]{Chris Simpson$^1$\thanks{E-mail:
    C.J.Simpson@ljmu.ac.uk}, Daniel Mortlock$^{2,3}$, 
Stephen Warren$^2$, 
Sebastiano Cantalupo$^4$, \newauthor Paul Hewett$^5$,
Ross McLure$^6$, Richard McMahon$^5$, and Bram Venemans$^7$\\
$^1$Astrophysics Research Institute, Liverpool John Moores University,
Liverpool Science Park, 146 Brownlow Hill, Liverpool L3 5RF\\
$^2$Astrophysics Group, Imperial College London, Blackett Laboratory,
Prince Consort Road, London SW7 2AZ\\
$^3$Department of Mathematics, Imperial College London, Blackett Laboratory,
Prince Consort Road, London SW7 2AZ\\
$^4$Department of Astronomy and Astrophysics, UCO/Lick Observatory,
University of California, 1156 High Street, Santa Cruz, CA 95064, USA\\
$^5$Institute of Astronomy, University of Cambridge, Madingley Road,
Cambridge CB3 0HA\\
$^6$Institute for Astronomy, Royal Observatory, University of
Edinburgh, Blackford Hill, Edinburgh EH9 3HJ\\
$^7$Max Planck Institut f\"{u}r Astronomie, K\"{o}nigstuhl 17, D-69117
Heidelberg, Germany}
\begin{document}

\date{Draft of \today}

\pagerange{\pageref{firstpage}--\pageref{lastpage}} \pubyear{2014}

\maketitle

\label{firstpage}

\begin{abstract}
  We present optical and near-infrared imaging of the field of the
  $z=7.0842$ quasar ULAS~J112001.48+064124.3 taken with the
  \textit{Hubble Space Telescope\/}. We use these data to search for
  galaxies that may be physically associated with the quasar, using
  the Lyman break technique, and find three such objects,
  although the detection of one in \textit{Spitzer Space
      Telescope\/} imaging strongly suggests it lies at $z\sim2$. This
  is consistent with the field luminosity function and indicates that
  there is no excess of $>L^*$ galaxies within 1\,Mpc of the quasar. A
  detection of the quasar shortward of the Ly$\alpha$ line is
  consistent with the previously observed evolution of the
  intergalactic medium at $z>5.5$.
\end{abstract}

\begin{keywords}
  dark ages, reionization, first stars --- galaxies: active ---
  galaxies: formation --- galaxies: high redshift --- quasars:
  individual: ULAS~J1120+0641
\end{keywords}

\section{Introduction}

Distant quasars at $z\ga6$ are uniquely able to facilitate studies of
the Universe in the first billion years after the Big Bang in two
ways. First, since they are the most luminous non-transient objects,
it is possible to measure the opacity of the intergalactic medium
(IGM) along the line-of-sight due to absorption of photons by neutral
hydrogen and probe the epoch of reionization. Second, by virtue of the
fact that they already contain $\sim10^9\,M_\odot$ black holes, it is
believed that quasars are located in the most overdense regions of the
early Universe (e.g., Springel et al.\ 2005; Sijacki, Springel, \&
Haehnelt 2009; Costa et al.\ 2014; but see Fanidakis et al.\ 2013 for
an alternative view). They should therefore act as beacons to
(proto-)clusters of high-redshift galaxies and permit much more
efficient observations of these galaxies.

Spectroscopy of quasars shortward of the redshifted Ly$\alpha$
emission line has shown a rapid increase in the optical depth from
absorption at $z_{\rm abs}>5.5$ (Fan et al.\ 2006), indicating that
$z\sim6$ is the end of the epoch of reionization. When compared to the
Thomson optical depth to microwave background photons, which is
consistent with instantaneous reionization at $z=11.1\pm1.1$ (Planck
Collaboration XVI 2014), this implies that reionization was an
extended and/or location-dependent process. However, due to the large
cross-sections for absorption, rest-frame far-ultraviolet spectroscopy
struggles to measure the neutral hydrogen fraction, $x_{\rm H\,I}$, if
$x_{\rm H\,I} \ga 10^{-3}$ and so has little hope of measuring its
evolution. The 21-cm hyperfine transition of H{\sc~i} has a much lower
cross-section and therefore has diagnostic power at higher neutral
fractions (Carilli, Gnedin, \& Owen 2002) but it requires a bright
radio source to be found at $z\ga6$ -- and this has yet to happen.

The quasar ULAS~J112001.48+064124.3 (hereafter ULAS~J1120+0641;
Mortlock et al.\ 2011) provided a new method of measuring the
neutrality of the IGM. Discovered with a combination of infrared
imaging from the UKIDSS Large Area Survey (Lawrence et al.\ 2007) and
optical imaging from the Sloan Digital Sky Survey (SDSS; Abazajian et
al.\ 2009), augmented with deeper imaging from the Liverpool
Telescope, its redshift was originally measured as $z=7.085\pm0.003$
from the broad ultraviolet emission lines, later refined to $z=7.0842$
from a detection of the [C{\sc~ii}]~$\lambda$158\,$\mu$m line by
Venemans et al.\ (2012). While the lack of a continuum detection
shortward of the Ly$\alpha$ emission line does not place any stringent
constraints on the evolution of $x_{\rm H\,I}$ beyond the redshifts
studied with SDSS quasars (Fan et al.\ 2006), the spectrum around the
emission line displays a damping wing that implies $x_{\rm H\,I} >
0.1$, and a remarkably small ionized near zone of $\sim2$\,Mpc in
extent (Mortlock et al.\ 2011; Bolton et al.\ 2011).  ULAS~J1120+0641
is therefore the first quasar to be discovered within the Epoch of
Reionization.

The environment of ULAS~J1120+0641 may therefore provide an
opportunity to study galaxy formation and evolution in a currently
unique environment, since distant active galaxies have often been
successfully used as signposts to locate overdensities in the
Universe. Searches for Lyman$\alpha$ emitters (LAEs) and/or Lyman
break galaxies (LBGs) have found excess numbers of objects around
radio galaxies at $z=4.11$ (TN~J1338$-$1942; Venemans et al.\ 2002)
and $z=5.19$ (TN~J0924$-$2201; Venemans et al.\ 2004; Overzier et
al.\ 2006), and quasars at $z=5.82$ (SDSS~J0836+0054; Zheng et
al.\ 2006) and $z=6.28$ (SDSS~J1030+0524; Stiavelli et al.\ 2005). In
an attempt to make a more reliable statistical determination of the
environments of distant quasars, Kim et al.\ (2009) studied
\textit{Hubble Space Telescope\/} (\textit{HST\/}) imaging of the
fields of the five most distant quasars known (at the time of their
proposal), finding overdensities of high-redshift LBG candidates in
two fields (including SDSS~J1030+0524), and underdensities in two
fields, with one field having the same sky density as seen in the
Great Observatories Origins Deep Survey (GOODS). There was no clear
interpretation of these results, with the underdense fields being
particularly surprising since the majority of the survey volume in
each field lies at a large radial distance from the quasar and so
should not be affected.

\begin{figure}
\resizebox{\hsize}{!}{\includegraphics{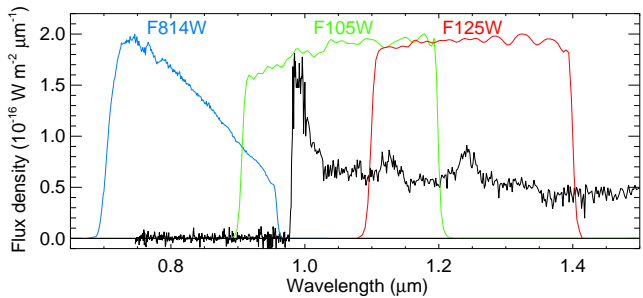}}
\caption[]{Spectrum of ULAS~J1120+0641 (Mortlock et al.\ 2011),
  overlaid with the \textit{HST\/} filters used in this work.}
\label{fig:qsolbg}
\end{figure}

Using semi-analytical prescriptions applied to the Millennium Run dark
matter simulation (Springel et al.\ 2005), Overzier et al.\ (2009)
find that there is sufficient scatter in the relationship between halo
mass and the number of star-forming galaxies in the vicinity that even
very massive haloes may not show an overdensity of galaxies. However,
they do not discuss whether quasars might represent a biased, rather
than a random, subset of massive haloes. The same authors note that
even when there is an overdensity of galaxies, the small field of view
of the \textit{HST\/} Advanced Camera for Surveys may fail to reveal
it. Larger fields of view are possible from the ground, but the
achievable depths are brighter due to the much higher sky
background. While Ba\~{n}ados et al.\ (2013) found no excess of either
LBGs or LAEs around the $z=5.72$ quasar ULAS~J0203+0012 (chosen
because Ly$\alpha$ at this redshift falls between the bright sky
lines), Utsumi et al.\ (2010) identified an excess of LBGs around the
then-most-distant known quasar, CFHQS~J2329$-$0301 at $z=6.43$. These
objects appeared to be located in a ring at a projected proper
distance of 3\,Mpc, which is comparable to the size of the ionized
near zones around $z\sim6$ quasars (Carilli et al.\ 2010). However,
the luminosities of these galaxies suggest they reside in massive
subhaloes and would be unaffected by the quasar's ionizing radiation
(Kashikawa et al.\ 2007), so the radiation field could not explain the
absence of similar galaxies at smaller quasarcentric distances.

In this paper we present \textit{HST\/} imaging of the field around
ULAS~J1120+0641 in three filters, designed to identify Lyman break
galaxies associated with the quasar (Fig.~\ref{fig:qsolbg}). The
format of this paper is as follows. In Section~2 we describe the
\textit{HST\/} observations and reduction, and in Section~3 we explain
how we identify $z\sim7$ galaxies. Section~4 presents a discussion of
the environment of the quasar and constraints on the evolution of the
Gunn--Peterson (Gunn \& Peterson 1965) optical depth, while a summary
of our results is provided in Section~5. We adopt a $\Lambda$CDM
cosmology with $H_0 = 70$\,km\,s$^{-1}$\,Mpc$^{-1}$ and $\Omega_{\rm
  m} = 1-\Omega_\Lambda = 0.3$. All magnitudes are on the AB system
(Oke \& Gunn 1983).

\section{Observations and reduction}

\begin{figure}
\resizebox{\hsize}{!}{\includegraphics{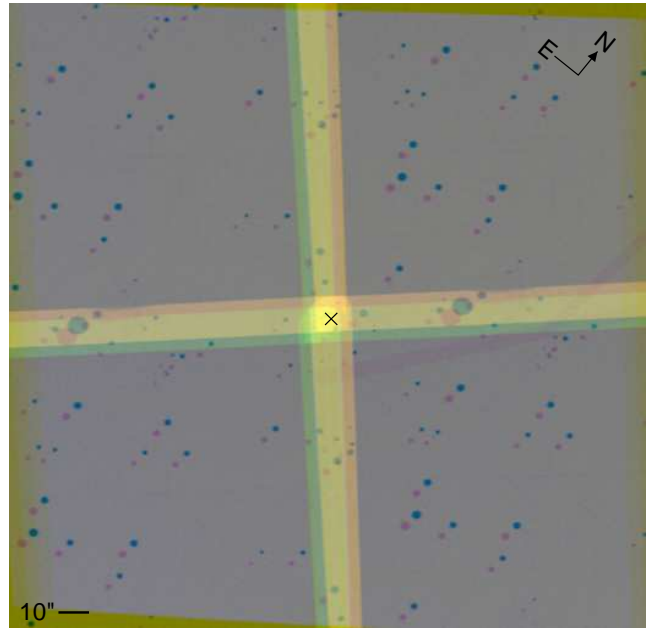}}
\caption[]{Exposure map of the field around ULAS~J1120+0641. The blue,
  green, and red channels represent the exposure times in the F814W,
  F105W, and F125W filters, respectively. A square-root scaling has
  been used to better represent the relative depths, and the circular
  `blobs' of dead pixels in the WFC3/IR detector are clearly
  visible. The cross indicates the location of the quasar, and a
  10-arcsec scale bar is shown for scale.}
\label{fig:exp_map}
\end{figure}

The field of ULAS~J1120+0641 was observed with the Advanced Camera for
Surveys (ACS) and the infrared channel of the Wide-Field Camera 3
(WFC3/IR) on board \textit{HST\/}, for programme GO-13039. Due to the
low ecliptic latitude of the target, all observations were performed
with the LOW-SKY constraint, to reduce the sky background.

The ACS imaging was performed using the F814W filter and the Wide
Field Camera (WFC) and comprised 13 orbits, spread across five
visits. Four of these visits had their pointing centres sequentially
offset by approximately 4\,arcsec to cover the gap between the two WFC
CCDs, and each visit consisted of six dithered exposures, two per
orbit. A single exposure was taken in the final visit, to give a total
exposure time of 28\,448\,s.

The WFC3/IR imaging was undertaken as a 2$\times$2 mosaic across four
visits. Within each visit, eight exposures in the F125W filter
(totalling 2116\,s) were taken in the first orbit, while a further
eight exposures in F105W (totalling 4416\,s) were split across the
next two orbits. Each set of eight exposures was made in two small box
dithers, separated by an offset large enough to straddle the largest
of the detector's bad pixel `blobs'. Adjacent frames in the mosaic
overlapped by 10\,arcsec to faciliate accurate registration. This
resulted in the region around the quasar itself being observed for
four times longer than most of the field. Fig.~\ref{fig:exp_map} shows
the effective exposure map produced.

\begin{figure}
\resizebox{\hsize}{!}{\includegraphics{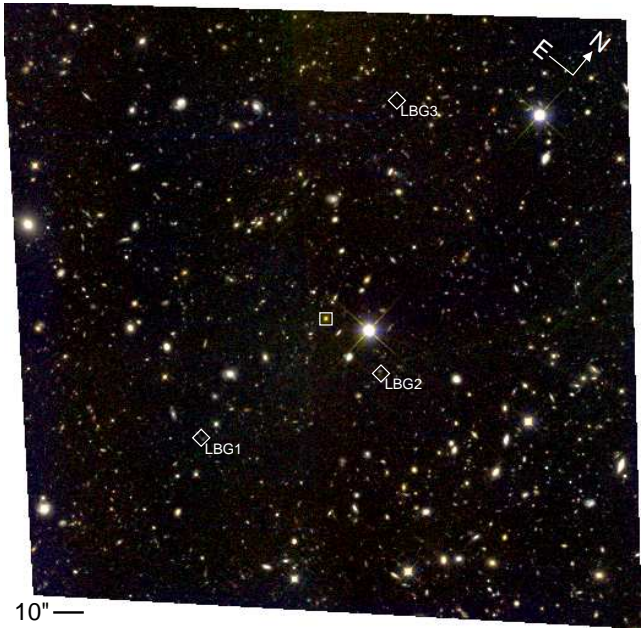}}
\caption[]{Combined \textit{iYJ\/} image of the field around
  ULAS~J1120+0641. The quasar is indicated with a white square, while
  the locations of the candidate Lyman break galaxies (see
  Section~\ref{sec:lbgs}) are marked with diamonds. The 10-arcsec
  scale bar represents 52\,kpc (proper) at the distance of the
  quasar.}
\label{fig:image}
\end{figure}

The pipeline-processed images were combined using the
\textit{astrodrizzle\/} task in PyRAF, initially using an output pixel
size equal to the input pixel size. The initial combination of the
WFC3/IR data revealed the presence of astrometric offsets between the
visits, due to different guide stars. These offsets were measured by
combining the F105W images in each visit separately and comparing the
coordinates of objects in the overlap regions. The world coordinate
systems in the image headers were then updated to correct for these
offsets, and new mosaics in F105W and F125W were made. We estimate the
uncertainty in these corrections to be less than 10\,mas by summing
the offsets around the four visits in the final mosaic, which should
total zero.

The combined, charge transfer efficiency-corrected ACS image revealed
no such offsets, but residual structure was seen in the background,
with a peak-to-peak variation of 1.5\,per cent of the sky level. We
investigated whether this was due to flatfielding errors by
constructing a corrective flatfield from our pipelined images. All 25
ACS images were scaled to the same sky level and this stack was median
filtered before the small-scale structure was removed by estimating
the variation in the background with SExtractor (Bertin \& Arnouts
1996). The individual images were then divided by a unity-normalized
version of this frame and the new images run through
\textit{astrodrizzle\/}, but no improvement was seen. We also
considered the possibility that the structure was due to the bias
level by median-filtering the images after normalizing the sky levels
by applying additive, rather than multiplicative, offsets but this had
a detrimental effect on the final drizzled image. This residual
structure was therefore removed from the final combined image using
the SExtractor-estimated background.

New versions of the two WFC3/IR mosaics were produced by drizzling
them onto images with the same pixel size (0.05\,arcsec) and world
coordinate system as the ACS image, to improve the resolution and aid
in the measurement of object colours. Object positions were then
compared in each of the three images and an offset of 90\,mas found
between the ACS and WFC3/IR images. This was corrected by altering the
headers of the ACS images and the stacked images were made a final
time, with the offsets between object positions in the different
filters found to be less than 20\,mas.

The region of sky imaged in all three filters measures
210$\times$225\,arcsec$^2$, corresponding to a projected linear
dimension of approximately 1.1\,Mpc (proper) at the redshift of
ULAS~J1120+0641. The drizzled WFC3/IR images both have a point spread
function FWHM of approximately 0\farcs21, while the ACS image has a
FWHM of $\sim$0\farcs12. A psf-matched version of the ACS image was
created by smoothing it with a Gaussian filter, and photometric
measurements were made from this smoothed image. We determine the
noise properties of our images by measuring the flux in 10\,000
apertures placed at random locations in each image and fitting a
Gaussian to the histogram of counts below +1$\sigma$. In the
0.5-arcsecond diameter apertures within which we measure colours, a
signal-to-noise ratio of 3 corresponds to magnitudes of
$i_{814}=28.50$, $Y_{105}=27.96$, and $J_{125}=27.39$. The depth of
the F814W image is limited by a combination of artifacts from the
charge transfer inefficiency not fully removed by the pipeline and
structure in the background on scales comparable to the extents of the
brightest objects that could not be removed. A colour image made from
the three individual filters is shown in Fig.~\ref{fig:image}.

\section{Analysis}

\begin{figure}
\resizebox{\hsize}{!}{\includegraphics{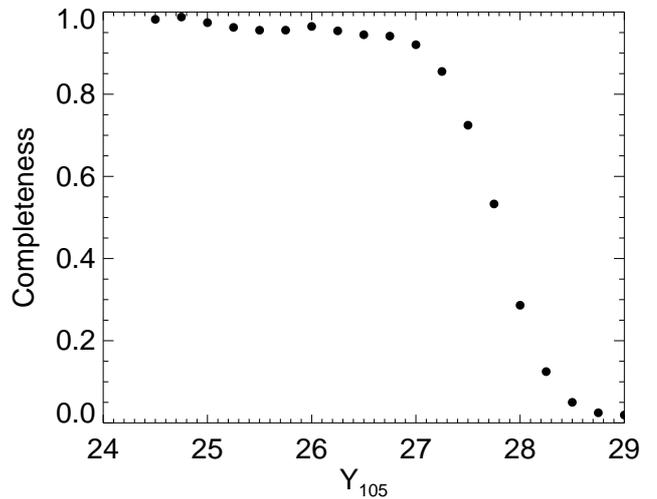}}
\caption[]{Completeness for $z\approx7$ galaxies as a function of
  total $Y_{105}$ magnitude, assuming galaxies have
  $Y_{105}-J_{125}=0.23$. Under the assumption that
  $Y_{105}-J_{125}=0.0$, all the points should be moved 0.07\,mag
  brighter.}
\label{fig:completeness}
\end{figure}

\begin{figure}
\hspace*{0.075\hsize}\resizebox{0.85\hsize}{!}{\includegraphics{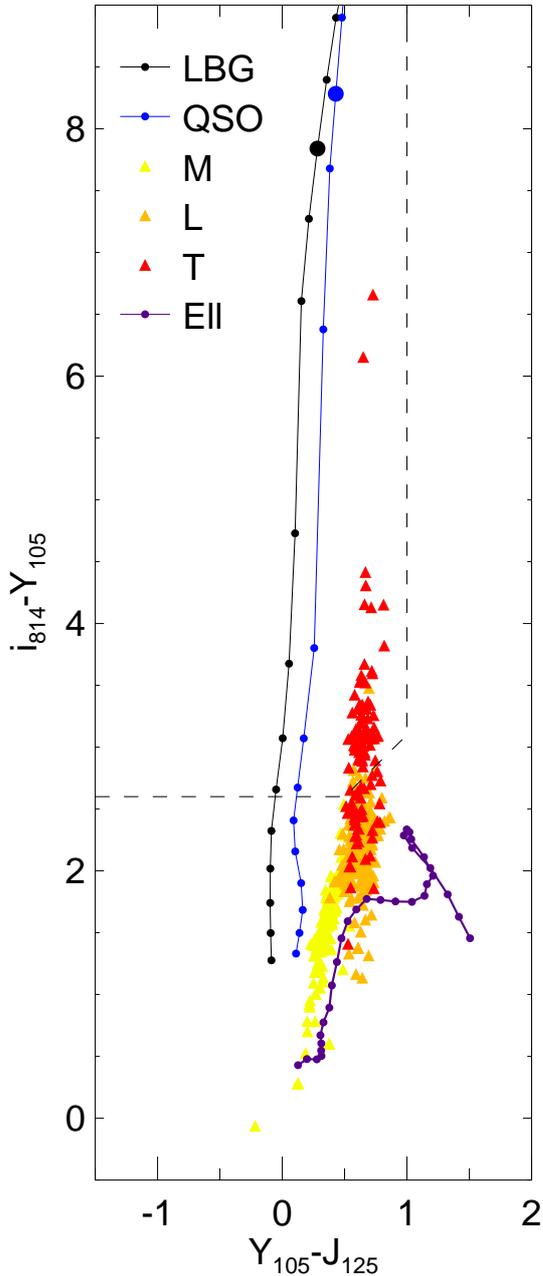}}
\caption[]{Colour--colour diagram used in the selection of LBGs at
  $z\sim7$. The loci for $z\geq6$ galaxies and quasars are shown by the
  black and blue lines, with points marked every $\Delta z=0.1$ and
  large symbols at $z=7.1$ (corresponding to the redshift of
  ULAS~J1120+0641). The purple line shows the locus of a non-evolving
  elliptical galaxy over the range $0\leq z\leq3$, which only
  approaches our selection box at high redshift, $z\sim2$, when any
  such galaxies will be younger and hence bluer. The locations of
  stars and brown dwarfs are shown by the filled triangles.}
\label{fig:iyj_model}
\end{figure}

\subsection{Object detection and photometry}

We follow the method of Szalay, Connolly, \& Szokoly (1999) to detect
objects, using a combined noise-weighted F105W+F125W image, referred
to as the $R$ image. SExtractor is used to find groups of contiguous
pixels with values above the threshold $R=3.35$ where the
probabilities of misclassifying sky and object pixels are equal. By
counting the number of objects detected with negative fluxes in one or
both filters, we can infer the level of contamination from spurious
detections for different minimum size thresholds, and adopt a value of
14 as producing a catalogue that is 97\,per cent genuine, based on the
number of detected sources with negative fluxes. Magnitudes are
measured in circular apertures of 0.5\,arcsec diameter as well as
elliptical apertures of 2.5 Kron (1980) radii, with a 0.1\,mag
correction applied in the manner of Oesch et al.\ (2010a) to account
for light outside this aperture. We exclude objects detected in
regions of sky where the exposure time in any filter is less than
one-third of the planned exposure times listed in the previous
section, and calculate an effective survey area of 11.1\,arcmin$^2$.

We estimate the completeness of our catalogue by inserting artificial
galaxies into the $R$ image and finding how many we recover. The
galaxies have half-light radii drawn from a normal distribution with a
mean of 0.7\,kpc and a standard deviation of 0.3\,kpc, following Oesch
et al.\ (2010b), with a minimum size of 0.2\,kpc. Each galaxy has a
S\'ersic index drawn from a uniform distribution between 1.0 and 4.0
and is convolved with the point spread function measured from an
isolated star. Since the F105W image is deeper, and hence more heavily
weighted in the construction of the $R$ image, the completeness is
most sensitive to $Y_{105}$ magnitude, with a secondary colour term.
Assuming a colour of $Y_{105}-J_{125}=0.23$ (see below), we obtain the
points shown in Fig.~\ref{fig:completeness}, with 50\,per cent
completess at $Y_{105}=27.78$ (approximately $M^*+0.4$ according to
the $z\sim7$ luminosity function of McLure et al.\ 2013). A colour of
$Y_{105}-J_{125}=0.0$ would cause the points to be shifted 0.07\,mag
brighter. The maximum completeness is 96\,per cent, indicating the
fraction of the image that is free from bright stars and galaxies.

\subsection{Lyman break galaxies at z$\sim$7}
\label{sec:lbgs}

\begin{figure}
\hspace*{0.075\hsize}\resizebox{0.85\hsize}{!}{\includegraphics{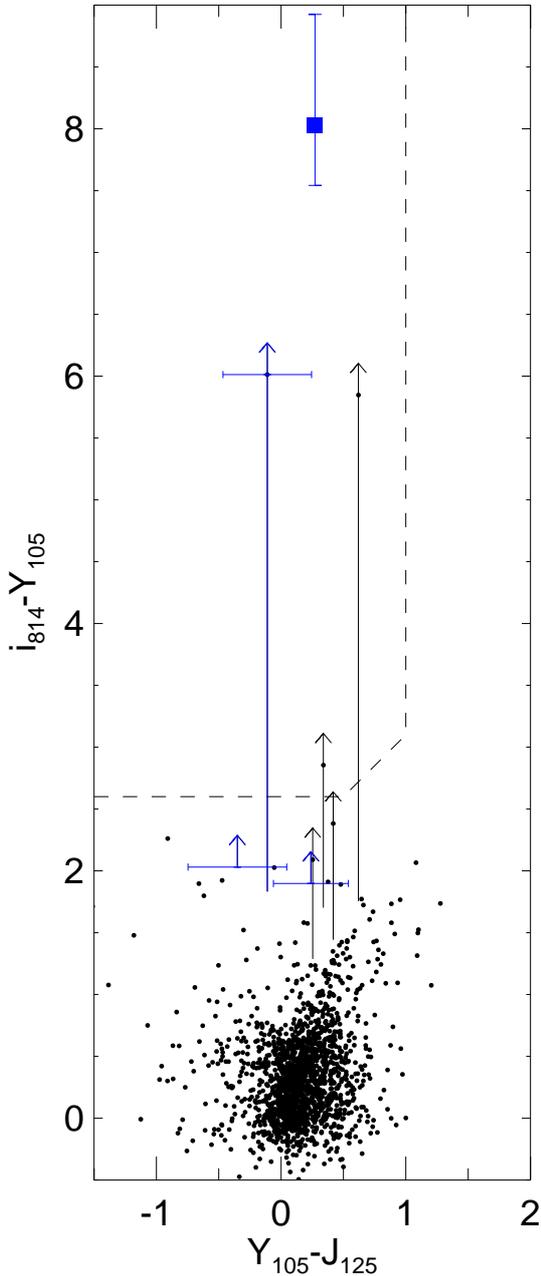}}
\caption[]{Colour--colour diagram for sources detected in the field of
  ULAS~J1120+0641. Objects selected as LBGs are in blue (the quasar
  itself is shown by a filled square) with 1$\sigma$ error bars
  plotted. If the 90\% confidence interval on the F814W flux of an
  object (calculated as described in the text) includes zero, a
  vertical line with an upward-pointing arrow is drawn to show the
  extent of this confidence interval.  The dashed line indicates our
  selection critera for $z\sim7$ objects, as in
  Fig.~\ref{fig:iyj_model}. Only objects with a signal-to-noise ratio
  $>5$ in $Y_{105}$ are plotted, although this criterion is not used
  in the selection of the LBG candidates.}
\label{fig:colcol}
\end{figure}

We model the colours of stars and galaxies through our three filters,
using the \textsc{synphot} package available in \textsc{iraf}, to
determine the appropriate selection criteria for LBGs at $z\sim7$.
LBGs are modelled as 100-Myr old bursts with a constant star-formation
rate, using Bruzual \& Charlot (2003) models, while quasar colours are
determined using the composite spectrum of Vanden Berk et al.\ (2001).
In both cases, attenuation from the the IGM was modelled in the manner
of Becker, Rauch, \& Sargent (2007). Colours of a non-evolving
elliptical galaxy are computed from the data of Coleman, Wu, \&
Weedman (1980), while the colours of late-type stars and brown dwarfs
are derived from the SpeX Prism Spectral Libraries.

Our selection criteria for LBGs, based on Fig.~\ref{fig:iyj_model}
are:
\begin{eqnarray}
i_{814} - Y_{105} & > & 2.6 \nonumber \\
Y_{105} - J_{125} & < & 1.0 \nonumber \\
i_{814} - Y_{105} & > & Y_{105} - J_{125} + 2.1
\end{eqnarray}

Other \textit{HST\/} searches for $z\ga7$ LBGs have taken advantage of
extant ultra-deep optical imaging in several bluer filters where
\textit{bona fide\/} LBGs will have zero flux, and which can therefore
be used to reduce the number of contaminants. This enables the use of
less robust colour--colour diagrams, with a smaller separation between
genuine $z\sim7$ LBGs and the key contaminants, which are $z\sim2$
galaxies, and less depth in the `dropout' filter. Since we lack such
ancillary imaging, a robust sample is required from our three-colour
imaging alone, and this is the reason for using F814W rather than
F850LP (e.g., compare Fig.~\ref{fig:iyj_model} to fig.~1 of Oesch et
al.\ 2010a). Unfortunately, this precludes a standard method of
analysis, as a limit of $z_{814}>30.0$ would need to be reached to
place a $Y^*_{105}=27.4$ galaxy within our selection region. Instead,
we adopt a Bayesian approach to dealing with upper limits in our F814W
data. If an object has a measured flux $\hat{F}$, then the probability
distribution for its true flux $F$ is given by
\begin{equation}
P(F|\hat{F}) \propto P(\hat{F}|F) P(F) \, .
\end{equation}
If we assume that the noise is Gaussian with a standard deviation
$\sigma$, which is usually true for faint objects where the
observations are background-limited, then
\begin{equation}
P(\hat{F}|F) = \exp (-[\hat{F}-F]^2/2\sigma^2)
\end{equation}
and so, assuming a uniform prior for $F\geq0$, we can infer the
probability that the true flux of an object is fainter than some limit
$F_{\rm lim}$,
\begin{equation}
P(F<F_{\rm lim}|\hat{F}) = 1 - 
\frac{{\rm erfc}([F_{\rm lim}-\hat{F}]/\sqrt{2}\sigma)}%
{{\rm erfc}(-\hat{F}/\sqrt{2}\sigma)}
\end{equation}
where erfc is the complementary error function. We extend this
analysis by considering the uncertainty in the $Y_{105}$ measurement,
which we assume to be Gaussian, since the limiting flux $F_{\rm lim}$,
in $i_{814}$, that we wish to consider is a function of the true
$Y_{105}$ flux (Equation~1).

\begin{figure}
\resizebox{\hsize}{!}{\includegraphics{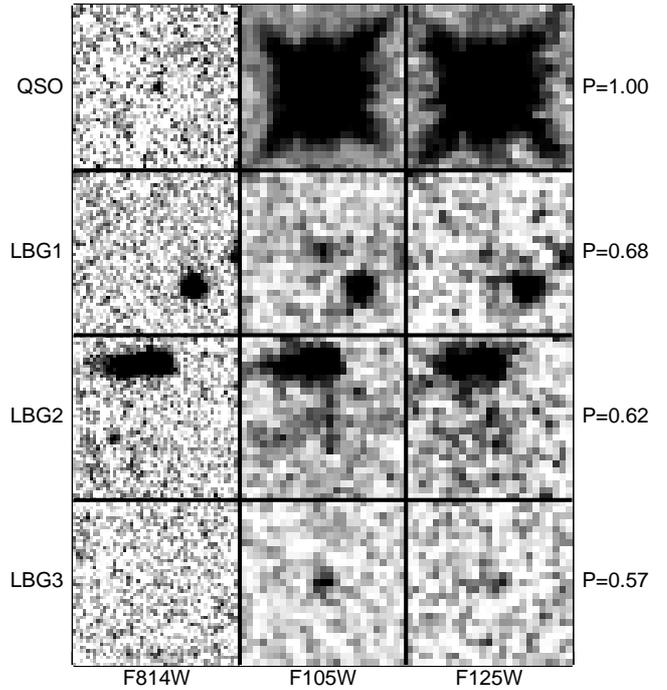}}
\caption[]{Images (2.5$"$ on a side) of the four objects listed in
  Table~\ref{tab:photdata}. The orientation is the same as
  Fig.~\ref{fig:image}, and the WFC3/IR images have been binned
  2$\times$2. The Bayesian probabilities that these have
  $i_{814}-Y_{105}>2.6$ are shown on the right-hand side of the
  figure.}
\label{fig:postages}
\end{figure}

\begin{table*}
  \caption[]{Photometry of candidate $z\sim7$ objects selected via the
    Lyman break technique. The $Y_{105}$ magnitude is a total (Kron)
    magnitude, while the signal-to-noise ratios and colours are
    measured in a 0.5-arcsecond diameter aperture. Statistical
    1$\sigma$ uncertainties are given, or upper limits at 90\%
    confidence.}
\label{tab:photdata}
\begin{tabular}{ccccrcrrrcc}
ID & RA & dec & $Y_{105}$ & $Y_{105}-J_{125}$ & $i_{814}-Y_{105}$ & 
SNR$_i$ & SNR$_Y$ & SNR$_J$ & $M_{1500}$ & $P$
\\ \hline
 QSO & 11:20:01.47 & +06:41:24.0 & 20.73$\pm$0.00 &  0.27$\pm$0.00 & 
    8.03$^{+0.95}_{-0.50}$ & 1.8 & 1761.7 & 1335.3 & $-$26.51$\pm$0.00 & 1.00 \\
 LBG1 & 11:20:01.93 & +06:40:27.9 & 26.34$\pm$0.23 & $-$0.35$\pm$0.40 & 
       $>$2.03 & $-$0.3 & 5.8 & 2.5 & $-$20.39$\pm$0.31 & 0.68 \\
 LBG2 & 11:19:59.77 & +06:41:21.2 & 25.78$\pm$0.23 &  0.24$\pm$0.30 & 
       $>$1.90 & $-$0.2 & 5.3 & 3.9 & $-$21.43$\pm$0.21 & 0.62 \\
 LBG3 & 11:20:03.30 & +06:42:34.4 & 27.23$\pm$0.48 & $-$0.11$\pm$0.36 & 
       $>$1.83 & 0.0 & 5.5 & 2.9 & $-$19.70$\pm$0.28 & 0.57
\\ \hline
\end{tabular}
\end{table*}

In order to determine an appropriate probability above which we should
consider a candidate drop-out galaxy to be plausible, we investigate
how our selection criteria fare when applied to the Hubble Ultra Deep
Field (HUDF). Oesch et al.\ (2010a) used WFC3/IR imaging in F105W and
F125W plus a deep ACS F850LP image to find $z\sim7$ galaxies, but the
HUDF F814W image (Illingworth et al.\ 2013) is not as deep as the
F850LP image and so we cannot make use of the full depth of the
WFC3/IR data. Instead we set a minimum $Y_{105}$ flux limit that is
0.4\,mag deeper than the 50\,per cent completeness limit in our data,
as is the extra depth of the HUDF F814W data compared to ours (this is
consistent with the factor of $\sim$2 longer exposure time). We
identify five candidate LBGs with $P\geq0.5$, including the four
brightest (in $Y_{105}$) sources from Oesch et al.\ (2010a). The
remaining source, which we designate UDFi-41436012 following Oesch et
al.'s naming convention, has $Y_{105}=26.69\pm0.17$ and is clearly
detected in F850LP but is absent in the shorter wavelength
filters. McLure et al.\ (2013) determine a robust photometric redshift
of $z=6.45$ for this source so it fails to make their $z>6.5$ sample,
but it is a bona fide dropout. Relaxing the lower limit on $P$
introduces sources that are demonstrably not at $z>6$ due to their
presence in the bluer images, and we therefore adopt $P\geq0.50$ as
the criterion for including an object in our sample of $z\sim7$
candidates. This means that an object with zero measured flux in our
F814W image will only be included in our sample if its 0.5-arcsec
aperture magnitude is $Y_{105}<27.52$, corresponding to a total
magnitude of $Y_{105}\approx27.2$.

As a final test, we add noise to the HUDF images to make their
effective depths comparable to those of our quasar field and apply our
detection algorithm again, now simply requiring that the flux ratio
between the two filters does not exceed 5, i.e., that the signal in
the detection image does not arise from only one of the WFC3/IR
images. We identify only the two brightest sources from Oesch et al.\
(2010a), plus UDFi-41436012, over an effective area of
4.9\,arcmin$^2$. Applying an identical analysis to our data, we obtain
the colour--colour diagram shown in Fig.~\ref{fig:colcol}, where four
sources are detected with $P>0.5$, one of which is the quasar
itself. These are listed in Table~\ref{tab:photdata} and postage
stamps in each of the three filters are shown in
Fig.~\ref{fig:postages}. We calculate rest-frame absolute magnitudes
at 1500\,\AA\ assuming the objects lie at the same redshift as
ULAS~J1120+0641.

\section{Discussion}

\subsection{The environment of ULAS~J1120+0641}

\begin{table}
  \caption[]{Measured flux densities in \textit{Spitzer\/}/IRAC channels
    1 and 2 for the three LBG candidates. Uncertainties have been
    determined from the root-mean-squared deviation of the counts in
    randomly-placed apertures and are therefore the same for all sources.}
\label{tab:irac}
\begin{center}
\begin{tabular}{crrrr}
& LBG1 & LBG2 & LBG3 & \\ \hline
IRAC1 & 1.45 & 5.58 & $-$1.06 & $\pm$0.94\,$\mu$Jy \\
IRAC2 & 0.17 & 6.83 & $-$0.54 & $\pm$0.81\,$\mu$Jy \\
\hline
\end{tabular}
\end{center}
\end{table}

\begin{figure}
\resizebox{\hsize}{!}{\includegraphics{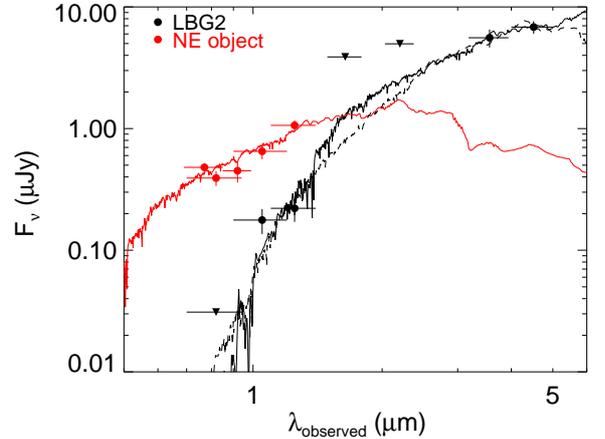}}
\caption[]{Spectral energy distributions of LBG2 (black points) and
  the galaxy to the northeast visible in Fig.~\ref{fig:postages} (red
  points), with best-fitting simple stellar populations at $z=2.49$
  and $z=0.27$, respectively, shown by solid lines. Upper limits are
  shown by downward-pointing triangles and include photometry from the
  deeper United Kingdom Infrared Telescope $H$ and $K$ imaging. The
  black dashed line shows an alternative fit to the LBG2 photometry
  with a much more heavily-reddened ($A_V=1.9$\,mag) populations at
  $z=1.5$.}
\label{fig:lbg2}
\end{figure}

Excluding the quasar itself, the objects we select are visually
extended and therefore cannot be cool brown dwarfs. The detection of
only three candidate $z\sim7$ galaxies in our field is surprising,
given that we detected the same number in the smaller (by a factor of
2.3) HUDF after making the data comparable in depth.  Clearly there is
no evidence for an overdensity of galaxies within 1\,Mpc of the
quasar, as might be expected if it is a signpost to a highly overdense
region of the early Universe. Conversely, the apparent underdensity
with respect to the HUDF cannot be due to the presence of the quasar
since the radial extent of our survey is $\sim250$\,Mpc, based on the
predicted redshift distribution given by our selection criteria and
the redshift-dependent parametrization of the field luminosity
function of Bouwens et al.\ (2011). Using this luminosity function,
and accounting for our selection function and incompleteness, we
predict 5.8 $z\sim7$ galaxies in our field (we predict 2.6 in the
HUDF), so there is a probability of 17\,per cent that we would find
three or fewer such galaxies.

Images of the ULAS~J1120+0641 field were taken with the Infrared Array
Camera (IRAC) on board the \textit{Spitzer Space Telescope\/} as part
of a programme to characterize the quasar's spectral energy
distribution. Inspection of these revealed a source offset by
$0.4\pm0.2$\,arcsec to the southwest of LBG2, i.e., away from the
bright galaxy seen at the top of the postage stamps in
Fig.~\ref{fig:postages}, but no detections at the locations of the
other LBG candidates (Table~\ref{tab:irac}). We attempted to determine
the nature of LBG2 and the nearby galaxy by fitting their spectral
energy distributions using the photometric redshift code EAzY
(Brammer, van Dokkum \& Coppi 2008). We assume that all the IRAC flux
arises from LBG2 and supplement the \textit{HST\/} photometry of the
nearby galaxy with data from Subaru/Suprime-Cam observations in the
$i$ and $z$ filters. Using simple stellar populations from Bruzual \&
Charlot (2003) attenuated by Pei's (1992) parametrization of the Small
Magellanic Cloud extinction law, we obtain photometric redshifts of
$z=2.49^{+0.14}_{-1.05}$ for LBG2 and $z=0.27^{+0.17}_{-0.18}$ for the
nearby galaxy. There is a strong anti-correlation between redshift and
extinction in the fit to LBG2, so lower redshift solutions require
more heavily-reddened populations (see Fig.~\ref{fig:lbg2}). All fits
to LBG2 predict $Y_{105}-J_{125}\approx1.1$, redder than observed, but
consistent with the data within 3$\sigma$. These fits suggest that the
IRAC photometry of LBG2 is contaminated by the nearby galaxy at a
level of only a few per cent, which does not significantly affect the
fit. We therefore conclude that LBG2 is most likely to be a galaxy at
$z\sim2$ and the probability of our finding only two or fewer $z\sim7$
galaxies in our field is 7\,per cent.

Although ULAS~J1120+0641 contains an extremely massive black hole that
would imply a massive halo, the dynamical mass estimated by Venemans
et al.\ (2012) from the [C{\sc~ii}]~$\lambda$158\,$\mu$m line is $<1.4
\times 10^{11}\,M_\odot$, much smaller than measured for local black
holes of similar mass. This supports the idea of Willott et al.\
(2005) that the large scatter between the mass of the black hole and
the mass of the halo in which it resides, coupled with the very steep
halo mass function, makes it likely that high-redshift quasars do not
sample the very massive haloes that their black hole masses would tend
to suggest.

Given its depth, our data cannot rule out an excess of sub-$L^*$
galaxies associated with the quasar. However, Stiavelli et al.\ (2005)
found an excess of galaxy candidates around SDSS~J1030+0524 despite
only reaching $z_{850}=26.5$, a full magnitude brighter than $M^*$ at
$z=6.28$, and our data would clearly be sensitive to an abundance of
galaxies that bright. It is therefore worth questioning whether the
presence of the quasar could have suppressed or delayed the formation
of galaxies in its immediate vicinity. This is a particularly
interesting question because the strong radiation field of the quasar
has an effect on the ionization state of the gas. It has been
suggested that the `missing satellites' problem can be explained if
baryons do not cool and form stars when their dark matter haloes are
below a certain mass at the time of reionization (Okamoto \& Frenk
2009). Since ULAS~J1120+0641 has reionized its immediate surroundings
earlier than the rest of the Universe, haloes that could have formed
stars elsewhere in the Universe may have been unable to do so around
the quasar.

Extrapolating the spectrum of Mortlock et al.\ (2011) as $S_\nu
\propto \nu^{-0.8}$, we calculate a luminosity at the Lyman limit of
$L_\nu = 1.3\times10^{24}$\,W\,Hz$^{-1}$, which would provide an
isotropic ultraviolet intensity of $J_\nu \approx 3.0 \times
10^{-23}$\,W\,m$^{-2}$\,Hz$^{-1}$\,sr$^{-1}$ at a radial distance of
550\,kpc (approximately the edge of our ACS field of view), or $J_{21}
\approx 30$ using the terminology of Kashikawa et al.\ (2007). Their
calculations suggest that haloes with virial masses $M_{\rm
  vir}\la10^{10}$M$_\odot$ could have their star formation
significantly delayed (by $\ga100$\,Myr) if exposed to a radiation
field of this strength for $\sim600$\,Myr (the time between $z=8.3$
and $z=4.87$), but more massive haloes would be unaffected. Since
ULAS~J1120+0641 is estimated to be accreting at close to the Eddington
limit (or even above it; Page et al.\ 2014), its luminosity in the
past must have been lower. Furthermore, the small size of the quasar's
near zone suggests that the ionizing radiation has only been able to
escape for a few Myr (Bolton et al.\ 2011), so even $M_{\rm
  vir}\approx10^{10}$M$_\odot$ haloes will have been unaffected. Since
an $L^*$ galaxy at $z\sim7$ has a stellar mass of $\sim10^9$M$_\odot$
(McLure et al.\ 2011), it is implausible that formation of galaxies
more massive than this could have been suppressed by the quasar.

We therefore consider the possibility that ULAS~J1120+0641 does not
reside in a massive halo. This is the scenario proposed by Fanidakis
et al.\ (2013), who argue, based on their semi-analytic modelling,
that the black holes in the most massive haloes have ended the phase
where they are accreting cold gas and are highly luminous, and have
already transitioned into the lower-luminosity feedback mode. This
would imply that the most massive haloes at $z=7.1$, where the galaxy
overdensities lie, contain black holes as massive as that in
ULAS~J1120+0641 that grew to this mass more rapidly before shutting
off. Given the apparent need for prolonged super-Eddington accretion
to explain how this black hole reached its mass, such a model requires
even more extreme accretion rates. It is worth noting that the first
$10^9$M$_\odot$ black holes in Fanidakis et al.'s (2012) model only
appear at $z<4$ and, as a result, the model fails to reproduce the
bright end of the observed quasar luminosity function at high
redshift. It is therefore of questionable merit to use these models to
predict the properties of the most extreme objects at $z>7$ and, while
the lack of any bright galaxy excess may be consistent with the claims
of Fanidakis et al.\ (2013), we note that our results are also
consistent with the work of Overzier et al.\ (2009).

\subsection{The opacity of the IGM}

Fig.~\ref{fig:postages} shows that the quasar is clearly detected in
the F814W image, which covers the observed wavelength range
$7000{\rm\,\AA} < \lambda < 9600\rm\,\AA$ ($866{\rm\,\AA} <
\lambda_{\rm rest} < 1188\rm\,\AA$) and samples wavelengths blueward
of those affected by the proximity effect of the quasar. The detection
of the quasar in this filter therefore gives some information on the
opacity of the IGM over the redshift range $4.76 < z < 6.89$. Although
the signal-to-noise ratio is only 1.8 in the 0.5-arcsecond diameter
aperture through which the photometry of Table~\ref{tab:photdata} is
measured, a 0.25-arcsecond diameter aperture on the unsmoothed image
gives a magnitude of $i_{814}=28.59\pm0.20$, after applying a 30\,per
cent correction to account for lost flux, as determined by TinyTIM
(Krist, Hook \& Stoehr 2011). An aperture of this size is warranted
because the position of the quasar is accurately determined from the
near-infrared imaging where it is bright. We nevertheless add a
10\,per cent flux uncertainty in quadrature, which accounts for a
positional uncertainty of up to 50\,mas.

We model the intrinsic quasar spectrum below Ly$\alpha$ using the
quasar spectrum presented by Mortlock et al.\ (2011), after first
scaling it to match the observed \textit{HST\/} photometry. Our
photometry is 14 (19) per cent fainter in the F105W (F125W) filter
than derived from the spectrum using the \textsc{synphot} package. As
the \textit{YJ\/} photometry from UKIRT agrees extremely well with
that derived from the spectrum we conclude that the fainter
\textit{HST\/} flux is due to variability between the
observations. Page et al.\ (2014) report a factor of approximately 2
decline in the X-ray flux over a 15-month period, and a 17\,per cent
decrease in the far-ultraviolet flux over a period of more than 2
years ($\sim100$\,days in the quasar rest-frame) is well within the
observed variability of similarly luminous quasars (Hook et al.\ 1994;
Welsh, Wheatley, \& Neil 2011). We extend the spectrum blueward of
Ly$\alpha$ in two ways. First, we follow Fan et al.\ (2006) in
assuming a spectrum $S_\nu \propto \nu^{-0.5}$ (i.e.,
$\alpha_\nu=-0.5$), which is consistent with the mean spectrum derived
from observations of the \textit{Far Ultraviolet Spectroscopic
  Explorer\/} (\textit{FUSE\/}) by Scott et al.\ (2004). Second, we
extrapolate the far-ultraviolet spectrum, which is well-fit by a
power-law with $\alpha_\nu=-0.8$, to $\lambda_{\rm
  rest}=1000$\,\AA\ and then use $\alpha_\nu=-1.4$ at shorter
wavelengths, based on \textit{HST\/}/Cosmic Origins Specotrograph
(COS) spectra (Shull, Stevans \& Danforth 2012).  Emission lines were
added with equivalent widths as given in table~3 of Scott et
al.\ (2004), and we assume that the IGM absorbs all flux at
$\lambda_{\rm rest}<912$\,\AA.

We model the IGM absorption following Fan et al.\ (2006), where the
effective Gunn--Peterson optical depth is given by
\begin{eqnarray}
\tau^{\rm eff}_{\rm GP} = 0.85 \left( \frac{1+z}{5} \right) ^{4.3} & & z\leq5.5 \\
\tau^{\rm eff}_{\rm GP} = 2.63 \left( \frac{1+z}{6.5} \right) ^\xi & & z>5.5
\end{eqnarray}
and determine the range of values of $\xi$ that are consistent with
our photometry. Fan et al.\ (2006) place a limit of
$\xi>10.9$ from their spectroscopic observations. We find that values
of $\xi = 10.8^{+0.6}_{-0.5}$ and $\xi = 10.6^{+0.6}_{-0.5}$ are
consistent with our two models.

We also account for the opacity of the IGM in the manner of Becker et
al.\ (2007), who model the optical depth distribution as a lognormal
distribution and find a single parametrization for its evolution
(without the break at $z\sim5.5$ required by Fan et al.\ 2006). This
model has no free parameters and predicts $i_{814}=28.64$ and
$i_{814}=28.74$ for the $\alpha_\nu=-0.5$ and $\alpha_\nu=-0.8$
models, respectively, both of which are consistent with our
measurement.

In both models, all the flux in the F814W filter is from wavelengths
just above the Ly$\beta$ absorption edge, where the overall opacity is
lowest, and therefore we are only sensitive to the opacity at
redshifts slightly larger than $z=(1+z_{\rm qso})\lambda_{\rm
  Ly\beta}/\lambda_{\rm Ly\alpha} - 1 = 5.82$. It is interesting to
note that this is where Mortlock et al.\ (2011) reported the detection
of flux in the quasar spectrum (an additional, less significant,
transmission spike at $z=6.68$ has subsequently been determined to not
be real). However, this feature has a flux of
$\sim6\times10^{-21}$\,W\,m$^{-2}$ (from a higher signal-to-noise
ratio X-Shooter spectrum; George Becker, private communication), which
is only one-half of that needed to explain the F814W flux. This
implies that there must be other regions along the line of sight where
the transmission is non-zero but too low to be identified in the
spectrum. A tighter constraint on the evolution of the opacity
requires ultra-deep imaging in a narrower filter shortward of the
quasar's Ly$\alpha$ emission.

\section{Summary}

We have presented new \textit{Hubble Space Telescope\/} observations
of the field around the $z=7.0842$ quasar ULAS~J1120+0641. We identify
three candidate $z\sim7$ galaxies from our \textit{HST\/} imaging but
conclude, on the basis of additional data, that only two are at this
redshift, with the third being a source at $z\sim2$. This is
consistent with the number of objects expected in a blank-field survey
of the same area and depth, and therefore conclude that there is no
excess of $\ga L^*$ galaxies within 1\,Mpc of the quasar. This cannot
be due to the influence of the quasar and we suggest that it is due to
the small field of view of our images, as proposed by Overzier et al.\
(2009). The detection of the quasar in the F814W filter provides a
constraint on the opacity of the IGM at $z\ga6$ and we find that the
model of Becker et al.\ (2007) successfully reproduces the observed
flux.

\section*{Acknowledgments}

SC acknowledges support from the NSF grant AST-1010004 and NASA
\textit{HST\/} grant GO-13033.06-A, RJM acknowledges ERC funding via
the award of a consolidator grant, and BV has been supported by the
ERC grant ``Cosmic Dawn.'' This paper is based on observations made
with the NASA/ESA \textit{Hubble Space Telescope\/}, obtained at the
Space Telescope Science Institute, which is operated by the
Association of Universities for Research in Astronomy, Inc., under
NASA contract NAS~5-26555. These observations are associated with
programme GO-13039, for which support was provided by NASA through a
grant from the Space Telescope Science Institute. This research has
benefitted from the SpeX Prism Spectral Libraries, maintained by Adam
Burgasser at http://pono.ucsd.edu/\~{}adam/browndwarfs/spexprism

\label{lastpage}

\end{document}